\documentclass[prl,twocolumn,letterpaper, superscriptaddress]{revtex4-1}
\usepackage{graphicx}
\usepackage{dcolumn}
\usepackage{bm}
\usepackage{color}
\usepackage{tabularx}
\usepackage{array}
\usepackage{amsmath}
\usepackage{stmaryrd}  

\begin{document}

\title{Strong magnon-photon coupling in ferromagnet-superconducting resonator thin-film devices}

\author{Yi Li}
\affiliation{Department of Physics, Oakland University, Rochester, MI 48309, USA}
\affiliation{Materials Science Division, Argonne National Laboratory, Argonne, IL 60439, USA}

\author{Tomas Polakovic}
\affiliation{Physics Division, Argonne National Laboratory, Argonne, IL 60439, USA}
\affiliation{Department of Physics, Drexel University, Philadelphia, PA 19104, USA}

\author{Yong-Lei Wang}
\affiliation{Materials Science Division, Argonne National Laboratory, Argonne, IL 60439, USA}
\affiliation{Research Institute of Superconductor Electronics, School of Electronic Science and Engineering, Nanjing University, 210093, Nanjing, China}

\author{Jing Xu}
\affiliation{Materials Science Division, Argonne National Laboratory, Argonne, IL 60439, USA}
\affiliation{Department of Physics, Northern Illinois University, Dekalb, IL 60115, USA}

\author{Sergi Lendinez}
\affiliation{Materials Science Division, Argonne National Laboratory, Argonne, IL 60439, USA}

\author{Zhizhi Zhang}
\affiliation{Materials Science Division, Argonne National Laboratory, Argonne, IL 60439, USA}
\affiliation{School of Optical and Electronic Information, Huazhong University of Science and Technology, Wuhan 430074, China}

\author{Junjia Ding}
\affiliation{Materials Science Division, Argonne National Laboratory, Argonne, IL 60439, USA}

\author{Trupti Khaire}
\affiliation{Materials Science Division, Argonne National Laboratory, Argonne, IL 60439, USA}

\author{Hilal Saglam}
\affiliation{Materials Science Division, Argonne National Laboratory, Argonne, IL 60439, USA}
\affiliation{Department of Physics, Illinois Institute of Technology, Chicago IL 60616, USA}

\author{Ralu Divan}
\affiliation{Center for Nanoscale Materials, Argonne National Laboratory, Argonne, IL 60439, USA}

\author{John Pearson}
\affiliation{Materials Science Division, Argonne National Laboratory, Argonne, IL 60439, USA}

\author{Wai-Kwong Kwok}
\affiliation{Materials Science Division, Argonne National Laboratory, Argonne, IL 60439, USA}

\author{Zhili Xiao}
\affiliation{Materials Science Division, Argonne National Laboratory, Argonne, IL 60439, USA}
\affiliation{Department of Physics, Northern Illinois University, Dekalb, IL 60115, USA}

\author{Valentine Novosad}
\email{novosad@anl.gov}
\affiliation{Materials Science Division, Argonne National Laboratory, Argonne, IL 60439, USA}

\author{Axel Hoffmann}
\email{hoffmann@anl.gov}
\affiliation{Materials Science Division, Argonne National Laboratory, Argonne, IL 60439, USA}

\author{Wei Zhang}
\email{weizhang@oakland.edu}
\affiliation{Department of Physics, Oakland University, Rochester, MI 48309, USA}
\affiliation{Materials Science Division, Argonne National Laboratory, Argonne, IL 60439, USA}

\date{\today}

\begin{abstract}

We demonstrate strong magnon-photon coupling of a thin-film permalloy device fabricated on a coplanar superconducting resonator. A coupling strength of 0.152 GHz and a cooperativity of 68 are found for a 30-nm-thick permalloy stripe. The coupling strength is tunable by rotating the biasing magnetic field or changing the volume of permalloy. We also observe an enhancement of magnon-photon coupling in the nonlinear regime of the superconducting resonator, which is mediated by the nucleation of dynamic flux vortices. Our results demonstrate a critical step towards future integrated hybrid systems for quantum magnonics and on-chip coherent information transfer.

\end{abstract}

\maketitle

Hybrid systems play a crucial role in quantum information processing \cite{Wallquist2009,XiangRMP2013,KurizkiPNAS2015}. In these systems, quantum states are conveyed from one platform to another by coherent coupling, which is represented by their mode hybridization \cite{BlaisPRA2004}. These coherent transduction will be necessary to utilize the advantage of different state variables \cite{WallraffNature2004,VerduPRL2009,SchusterPRL2010,KuboPRL2010,ZhuNature2011,VerhagenNature2012,ViennotScience2015}, such as microwave photons for quantum logic in superconducting circuitry and optical photons for long distance communication.

Recently, magnons have been considered as a new candidate of excitation for coherent information processing \cite{FlattePRL2010,HueblPRL2013,TabuchiPRL2014,ZhangPRL2014,GoryachevPRApplied2014,BhoiJAP2014,TabuchiScience2015,BaiPRL2015,ZhangnpjQuantumInfo2015,BauerPRB2015,LambertJAP2015,KostylevAPL2016,LachanceScienceAdvan2017,MorrisSREP2017,BoventerPRB2018}. Magnons are the collective excitation of exchange-coupled spins in ferromagnetic materials. They can conveniently couple to microwave photons via magnetic dipolar interaction. Especially, compared with paramagnetic spin ensembles which have been proposed as quantum memories \cite{ImamogluPRL2009,WesenbergPRL2009,SchusterPRL2010,KuboPRL2010,ProbstPRL2013}, ferromagnetic materials can provide much larger coupling strength and cooperativity, because they have spin densities 4 to 6 orders of magnitude higher than in diluted spins. This means magnons are capable of exchanging information with a much faster speed and for more cycles before losing coherency, while keeping small device dimensions. Coherent coupling between superconducting qubits and a single magnon has also been recently demonstrated \cite{TabuchiScience2015,LachanceScienceAdvan2017}, showing the potential for magnons to conduct real quantum operations. Furthermore, with new advances in spin-charge interconversion \cite{HoffmannIEEE2013,SinovaRMP2015}, the excitation of magnons in hybrid systems can be electrically detected via spin pumping \cite{MosendzPRL2010,BauerPRB2015,BaiPRL2015} and potentially other spin-transport phenomena \cite{LiuPRL2011,NakayamaPRL2013}.

Despite the progress in magnon-photon hybrid system, which are predominantly centered on yttrium iron garnet (YIG) ferrimagnets \cite{,HueblPRL2013,TabuchiPRL2014,ZhangPRL2014,GoryachevPRApplied2014,BhoiJAP2014,TabuchiScience2015,BaiPRL2015,ZhangnpjQuantumInfo2015,LambertJAP2015,KostylevAPL2016,LachanceScienceAdvan2017,MorrisSREP2017,BoventerPRB2018}, it is difficult to achieve on-chip integration because of the critical conditions in deposition and fabrication. In addition, the standard substrate for the growth of YIG thin films, Gd$_{3}$Ga$_{5}$O$_{12}$, experiences large magnetic losses at cryogenic temperatures \cite{CLiuPRB2018}, which will significantly reduce the coupling coherency for quantum applications. Thus it is desired to explore alternative magnetic systems for large-scale magnon-based hybrid quantum systems.

In this work, we demonstrate an all-on-chip circuit for magnon-photon hybrid system between a coplanar superconducting resonator and a permalloy (Ni$_{80}$Fe$_{20}$, Py) thin-film device. On the magnon side, Py is a classical metallic ferromagnet with well-known magnetic properties and industry-friendly deposition requirement. It exhibits five times larger spin density than YIG and allows even larger coupling strengths. On the photon side, a coplanar superconducting resonator has a much smaller mode volume than a macroscopic microwave cavity along with a higher quality factor, which allows more concentrated and long-lived photons to couple with magnons. We achieve a strong magnon-photon coupling strength of $g/2\pi=0.152$ GHz and cooperativity of $C=68$ for a small volume of Py ($V=400$ $\mu$m$^3$). By varying the dipolar coupling efficiency with the external magnetic field direction and changing the Py volume, the coupling strength $g$ can be also tuned in the hybrid system. Furthermore, we show that superconducting resonator can be easily driven into the nonlinear regime due to the dynamic magnetic flux vortices, which shows the quantum nature of the superconducting platform. Our results suggest the combination of superconducting resonator and metallic ferromagnets can be a promising platform for investigating on-chip quantum magnonics and spintronics, and brings new potential for coherent manipulation and long-distance propagation of spin information.

\begin{figure}[htb]
 \centering
 \includegraphics[width=3.2 in]{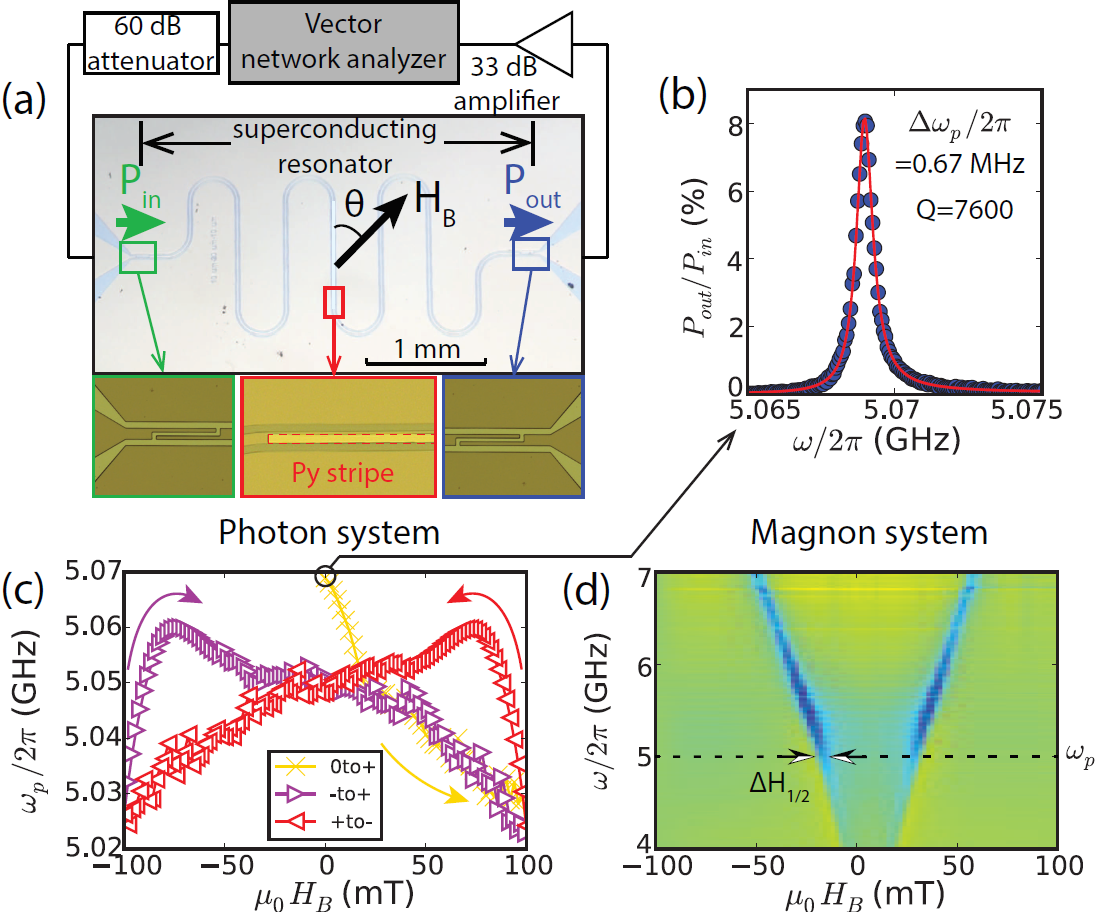}
 \caption{(a) The microwave circuit of a NbN superconducting resonator with a Py stripe. The green (blue) and red boxes show the capacitive coupling to the external circuit and the permalloy stripe, respectively. (b) Microwave power transmission of an unloaded superconducting resonator measured at $P_{in}=-55$ dBm after zero-field cooling. (c) Hysteresis evolution of $\omega_p$ for (b). (d) Ferromagnetic resonance spectra of a Py stripe measured at 1.4 K \cite{supplement}, with the linewidth at $\omega_p$ marked by arrows.}
 \label{fig1}
\end{figure}

Superconducting coplanar resonators were fabricated from 200-nm-thick NbN films by photolithography and reactive ion etching, Fig. \ref{fig1}(a). The NbN films were deposited on undoped Si substrates via ion-beam-assisted reactive sputtering technique at room temperature \cite{PolakovicAPLMaterials2018}. {\color{black} The resonator is capacitively coupled to the external microwave circuit at both the inlet and outlet \cite{WallraffJAP2008}}. Subsequently, a 30-nm Py thin-film stripe with lateral dimensions of $14\times900$ $\mu$m$^2$ is fabricated in close proximity and on top of the signal line of the resonator but electrically isolated from it by a 20-nm MgO insulating layer. The microwave response of the system is characterized by a vector network analyzer. A 60 dB attenuator is inserted before the microwave entering the circuit. Throughout the experiment the samples are cooled down to 1.4 K, which is well below the superconducting transition temperature of the NbN resonator, $T_c=14$ K.

\begin{figure*}[htb]
 \centering
 \includegraphics[width=5.5 in]{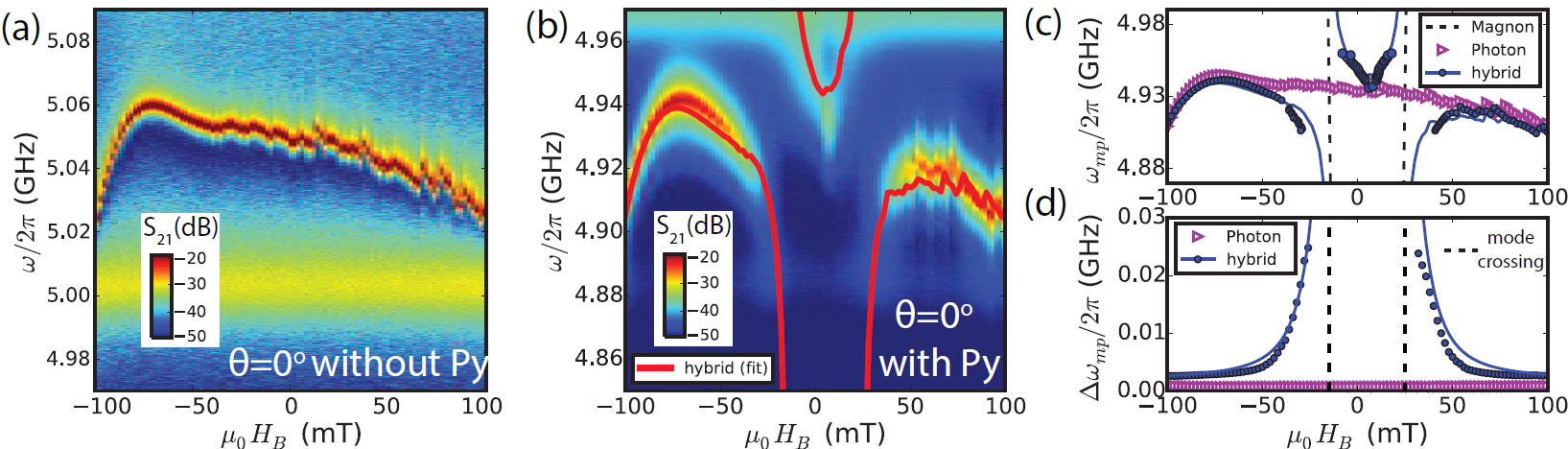}
 \caption{Characterization of a 30-nm Py stripe (L=900 $\mu$m) coupled to a NbN superconducting resonator, measured at $P_{in}=-55$ dBm and $\theta=0^\circ$. (a-b) Microwave transmission spectra $S_{21}=10\log(P_{out}/P_{in})$ of (a) the unloaded resonator and (b) the resonator loaded with the Py stripe. (c) Extracted $\omega_{mp}$ and (d) $\Delta\omega_{mp}$ from (a-b). In (c) the photon modes have been shifted by $-0.12$ GHz to match the hybrid modes. Dashed lines denote the magnon modes. Solid blue and red curves denote the fits.}
 \label{fig3}
\end{figure*}

The mode evolution of the individual magnon and photon systems are shown in Figs. \ref{fig1}(c) and (d), respectively, as a function of the in-plane magnetic field $\mu_0H_B$ along the stripline orientation ($\theta=0^\circ$, defined in Fig. \ref{fig1}a). For the photon subsystem (without the Py stripe), the NbN resonator exhibits a sharp peak at $\omega_p/2\pi=5.069$ GHz with a full-width-half-maximum linewidth $\Delta\omega_p/2\pi=0.67$ MHz, corresponding to a high quality-factor of $Q=7600$, see Fig. \ref{fig1}(b). The peak position corresponds to a dielectric constant, $\epsilon_r \sim 9.3$, similar to the value of 11.7 for Si. In addition, we obtain a hysteresis of $\omega_p$ by sweeping $\mu_0H_B$ (Fig. \ref{fig1}c). This behavior originates from the kinetic inductance variation from the nucleation of magnetic flux vortices in superconducting resonators \cite{BothnerPRB2012}. For the magnon subsystem, we have fabricated an individual Py stripe on a coplanar waveguide and measured its broad-band ferromagnetic resonance at 1.4 K \cite{supplement}. Two branches of resonance absorption are symmetrically located on the positive and negative fields, as shown in Fig. \ref{fig1}(d). The small field offset, $\sim$ 5 mT, comes from the hysteresis of the superconducting magnet coils. From Fig. \ref{fig1}(d), we can determine the linewidth of the ferromagnetic resonance as $\mu_0\Delta H_{1/2}=6.0$ mT at the frequency of $\omega_p$. This corresponds to a magnon damping rate of $\kappa_m/2\pi=(\gamma/2\pi)\mu_0\Delta H_{1/2}=0.178$ GHz, in which $\gamma/2\pi=(g_{eff}/2)\cdot28$ GHz/T is the gyromagnetic ratio and $g_{eff}=2.12$ is the g-factor for Py.

We then turn to the magnon-photon hybridization. Before the Py deposition, the superconducting resonator exhibits a continuous spectra when $\mu_0H_B$ is swept from $-100$ mT to $100$ mT, as shown in Fig. \ref{fig3}(a). After Py deposition, two avoided crossings appear symmetrically at positive and negative $\mu_0H_B$ (Fig. \ref{fig3}b). The mode anti-crossing indicates a strong coupling between the resonator photon and the ferromagnetic magnon. The extracted peak positions ($\omega_{mp}$) and linewidths ($\Delta\omega_{mp}$) of the spectra are summarized in Figs. \ref{fig3}(c) and (d), respectively. A frequency offset of $0.12$ GHz between Figs. \ref{fig3}(a) and (b) has been taken into account due to the local impedance change from the additional Py stripe. As shown in Fig. \ref{fig3}(c), the two anti-crossings are located where the two magnon branches of the Py stripe intersect with the photon mode of the resonator, which clearly indicates the strong magnon-photon coupling.

The transmitted power of the hybrid system can be expressed as \cite{HueblPRL2013,TabuchiPRL2014,ZhangPRL2014,BaiPRL2015}:
\begin{equation}\label{eq01}
{P_{out} \over P_{in}} = {\kappa_R \over i(\omega_p-\omega)+\kappa_p + {g^2 \over i(\omega_m-\omega)+\kappa_m}}
\end{equation}
where $\kappa_R$ is the capacitive coupling of the resonator to the external circuits and $g$ is the magnon-photon coupling strength. $\kappa_p$ is increased from Fig. \ref{fig1}(b) to $\kappa_p/2\pi=2.0$ MHz with the additional Py load, see Fig. \ref{fig3}(d) at high fields. Fig. \ref{fig3}(c) overlays the fits to the eigenmode solution of Eq. (\ref{eq01}) on top of the extracted $\omega_p$ from Fig. \ref{fig3}(b), with a single fit parameter $g/2\pi=0.152$ GHz \cite{supplement}. Note the large value of $g$ despite the small ferromagnetic volume with merely 30-nm of Py. To understand the origin, the coupling strength is expressed as $g=g_0\sqrt{N}$ where $N$ is the total number of spins and $g_0=\gamma \sqrt{\mu_0\hbar\omega_p/V_c}$ is the coupling strength of the superconducting resonator to a single Bohr magneton. Here $\hbar$ is the Planck constant, and $V_c$ is the mode volume of the resonator. Using the dimensions of the Py stripe and $\mu_0M_s=1$ T for the Py saturation magnetization, we calculate $N=3.25\times 10^{13}$ and $g_0=26.7$ Hz from the experiment. We highlight that our $g_0$ is three orders of magnitude larger compared with using a macroscopic cavity \cite{TabuchiPRL2014,ZhangPRL2014}. It comes from the small mode volume $V_c \sim 0.0051$ mm$^3$ for the coplanar resonator and indicates the significance of having a localized and concentrated photon mode volume to reach a strong coupling strength. In addition, compared with the similar superconducting resonator structure coupled to a YIG slab \cite{HueblPRL2013} ($g/2\pi=0.45$ GHz, $g_0/2\pi=2.5$ Hz and $N=4\times 10^{16}$), our $g_0$ is one order of magnitude larger because the nanoscale Py layer is in good proximity to the resonator and maintains optimal coupling efficiency. This yields a comparable $g$ of Py stripe but with three orders of magnitude less number of total spins than in the YIG slab. Despite the lossy Py, We still obtain a large cooperativity of the hybrid system as $C = g^2/\kappa_m\kappa_p=68$, which is a promising feature of coherent information exchange between photons and magnons in Py.

\begin{figure*}[htb]
 \centering
 \includegraphics[width=6.2 in]{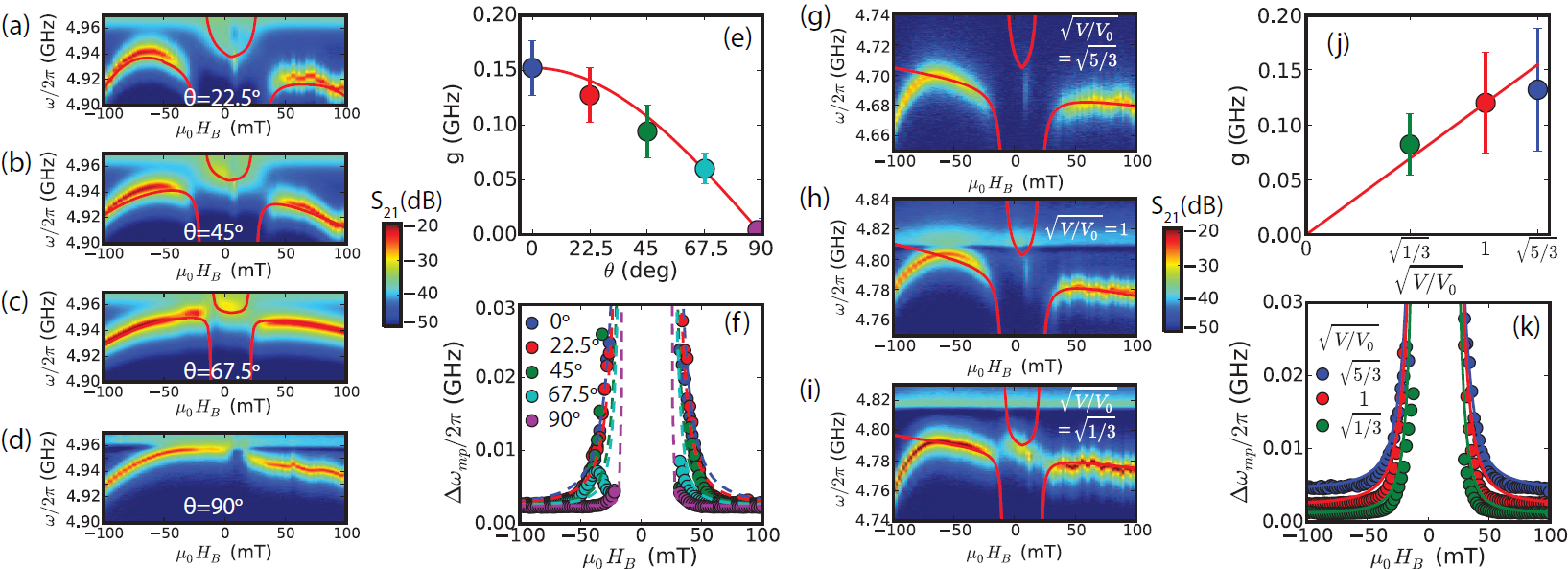}
 \caption{Tunable magnon-photon coupling. (a-d) Microwave transmission spectra of the NbN superconducting resonator loaded with a Py(30 nm) stripe with $L=900$ $\mu$m, from $\theta=22.5^\circ$ to $90^\circ$. (e) Extracted coupling strength $g$ as a function of $\theta$ with the fit. (f) $\Delta\omega_p$ as a function of $\mu_0H_B$ with the dashed fits. (g-i) Transmission spectra for different Py stripes: (g) $t=50$ nm, $L=900$ $\mu$m; (h) $t=30$ nm, $L=900$ $\mu$m; (i) $t=30$ nm, $L=300$ $\mu$m. (j) Extracted $g$ as a function of $\sqrt{V/V_0}$, where $V_0$ denotes the volume of Py(30 nm) stripe with $L=900$ $\mu$m. (k) $\Delta\omega_p$ as a function of $\mu_0H_B$, with the fitting curve also plotted.}
 \label{fig4}
\end{figure*}

In addition to the frequency shift, we also observe a linewidth shift for the hybrid modes \cite{BaiPRL2015}. In Fig. \ref{fig3}(d), when $\mu_0H_B$ is close to the anti-crossing regime, $\Delta\omega_{mp}$ quickly increases from the photon damping rate $\kappa_{p}/2\pi= 2$ MHz and approaches the magnon damping rate $\kappa_m$. This is due to the mixing of relaxation channels when the magnon and photon modes are hybridized, see the Supplemental Materials \cite{supplement}. We plot the theoretical prediction in Fig. \ref{fig3}(d) with the same input values of $g$, $\kappa_p$ and $\kappa_m$ in Fig. \ref{fig3}(c), and the linewidth of the hybrid modes can be reproduced. Between the two mode-crossing gaps (between -25 mT and +32 mT) the hybrid modes are influenced by the saturation state of the Py stripe, which significantly deviate from the macrospin model and are not shown.

The coupling strength $g$ is tunable by changing the dipolar coupling efficiency between magnons and photons as well as changing the total number of spins in Py. Figs. \ref{fig4}(a-d) show the microwave transmission spectra of the same device in Fig. \ref{fig3} at different $\theta$. As $\theta$ deviates from $0^\circ$, the mode anti-crossing becomes smaller and disappears at $90^\circ$. This is due to the change of dipolar coupling energy, $E=\mu_0M_\perp h_{rf}\cos\theta$, where the transverse components of the dynamic magnetization $M_\perp$ and microwave field $h_{rf}$ are no longer parallel and become orthogonal when $\theta=90^\circ$. The extracted $g$ can be modeled by a cosine function of $\theta$ (red curve) in Fig. \ref{fig4}(e). In addition, there are two additional observations in Figs. \ref{fig4}(a-d): the mode anti-crossing moves towards the lower biasing fields, and a spectral gap appears near $\mu_0H_B=0$ mT for $\theta=90^\circ$. They are due to the shape anisotropy of the Py stripe \cite{LiJAP2013}, which pins the Py magnetization along $\theta=0^\circ$ at low fields. To vary the total number of spins for magnons, we have fabricated a new series of Py stripes with different lengths ($L$) and thicknesses ($t$) and show their transmission spectra in Fig. \ref{fig4}(g-i). As the coupling strength is reduced, the mode hybridization becomes weaker and for the smallest Py volume in Fig. \ref{fig4}(i), the spectra go into the Purcell regime \cite{ZhangPRL2014} as $g$ becomes significantly smaller than $\kappa_m$. The extracted $g$, which are plotted in Fig. \ref{fig4}(j), follows a linear dependence on $\sqrt{V}$ within the errorbars.

\begin{figure}[b]
 \centering
 \includegraphics[width=3.2 in]{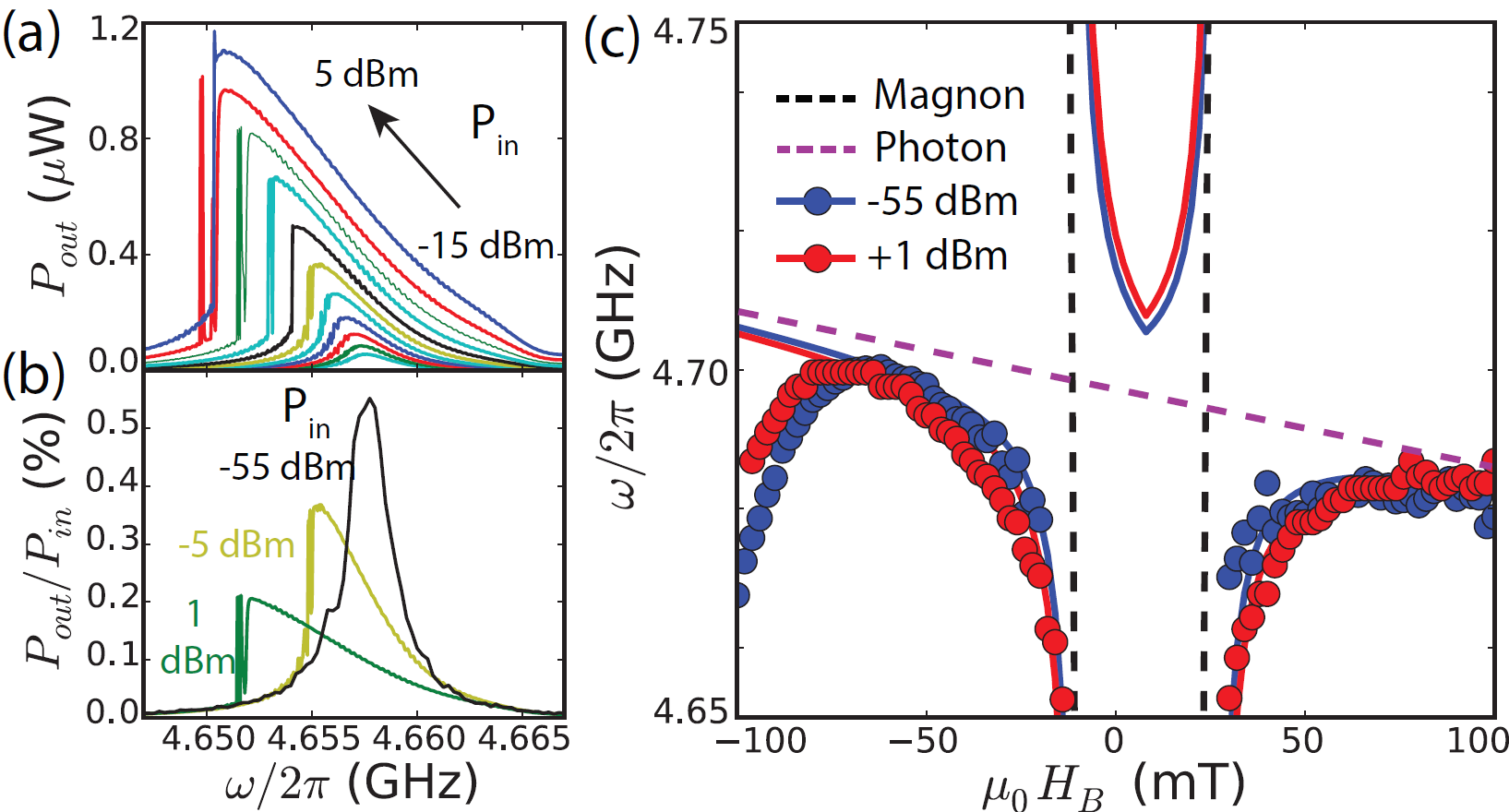}
 \caption{Magnon-photon coupling in the nonlinear regime. (a) Nonlinear resonance lineshapes from $P_{in}=-15$ dBm to $P_{in}=5$ dBm with a step of 2 dB for Py(50 nm) stripe with $L=900$ $\mu$m. (b) $P_{out}/P_{in}$ for $P_{in}=-55$ dBm, $-5$ dBm and 1 dBm. (c) Comparison of peak positions of the hybrid modes between $P_{in}=1$ dBm and $-55$ dBm.}
 \label{fig5}
\end{figure}

The evolution of $\Delta \omega_{mp}$ also changes accordingly for different magnon-photon coupling conditions. In Fig. \ref{fig4}(f) when the biasing field is far away from the anti-crossing regime ($\mu_0H_B=\pm 100$ mT), $\Delta \omega_{mp}$ shows a consistent value of $2.5$ MHz for the same Py device at different $\theta$. As $H_B$ approaches the anti-crossing conditions, $\Delta \omega_{mp}$ increases much slower for larger $\theta$, because the coupling strength is decreasing. This trend can be theoretically reproduced in dashed curves by taking different values of $g$ from Fig. \ref{fig4}(e) in the theoretical model \cite{supplement}. In Fig. \ref{fig4}(k), due to the variation of the dielectric loss from Py, the values of $\kappa_{p}$ are different, as 4.4 MHz, 2.3 MHz and 1.1 MHz for Figs. \ref{fig4}(g) to (i), respectively. By accounting for this $\kappa_p$ variation, the linewidths can be also well fitted for different Py volumes.

Next, we show that the photon mode in the hybrid system can easily go into the nonlinear regime, which is important for high-fidelity quantum operations \cite{YurkeJLT2008,MalletNPhys2009,OngPRL2012}. In Fig. \ref{fig5}(a), we show the output lineshapes of the superconducting resonator loaded with a Py(50 nm) stripe from $P_{in}=-15$ dBm to 5 dBm, at $\mu_0H_B=100$ mT and $\theta=0^\circ$. A nonlinear shift of the peak position towards the lower frequency is observed, with a critical power of $P_c=-5$ dBm for the lineshape to reach a vertical slope (Fig. \ref{fig5}b). This critical power is well below the typical threshold power for the magnon system alone to reach the nonlinear regime \cite{WangPRL2017}. The origin of the nonlinearity is the kinetic inductance variation of flux vortices \cite{GolosovskyPRB1995} in the NbN resonator, which leads to frequency downshifts as also observed in Fig. \ref{fig1}(c) at increasing $H_B$. We note that such vortex-induced nonlinearity in the superconducting resonator can be extended to the variation of a single vortex flux \cite{NsanzinezaPRL2014}, which shows potentials for conducting operation in the quantum limit. The dynamics of magnetic flux vortices also leads to an enhanced photon damping rate $\kappa_p$ \cite{PlourdePRB2009}, reflected by the reduction of maximal value of $P_{out}/P_{in}$ in Fig. \ref{fig5}(b).

Accompanied by the resonator nonlinearity, we also observe an enhanced magnon-photon coupling. In Fig. \ref{fig5}(c) we show the extracted peak positions of the hybrid mode for the 50-nm Py stripe. In the vicinity of the anti-crossing regimes ($\mu_0H_B$ close to the magnon branch), the peaks with $P_{in}=1$ dBm show a stronger mode repelling compared with $P_{in}=-55$ dBm. Fitting the data to Eq. (\ref{eq01}) yields a coupling strength of $0.158$ GHz at 1 dBm input, which is 14\% larger than the value of $0.139$ GHz at $-55$ dBm, in Fig. \ref{fig4}(j). This coupling enhancement is likely caused by the Meissner field trapping from the dynamic flux vortices \cite{TaupinNComm2016}, which will influence the distribution of the magnetic field at the superconducting stripline and thus change the dipolar coupling strength with the Py magnon system. Therefore, the magnon-photon coupling may be also used as an effective means for detecting flux vortex dynamics in hybrid superconducting devices.

In conclusion, we have demonstrated a new hybrid platform consisting of a superconducting resonator and a ferromagnetic metal, integrated on a Si substrate. We obtained a large magnon-photon coupling strength of 0.152 GHz and a cooperativity of 68 for a 30-nm-thick Py stripe. The coupling strength is adjustable by rotating the biasing field angle or changing the volume of Py. We also show that the superconducting resonator can easily reach the nonlinear regime, in which the efficiency of the magnon-photon coupling is improved. Our results indicate that the magnon-photon hybrid system is promising as a high-speed and coherent transducer for realizing circuit quantum electrodynamics \cite{LachanceScienceAdvan2017}, in microscopic magnetic devices that are compatible with on-chip designs. Furthermore, the magnon-photon hybrid system provides a means to transmit spin excitations coherently at long distance with photon excitations \cite{ZhangNComm2015,BaiPRL2017}, outperforming the currently-reported micro-meter propagation using pure spin currents \cite{ValenzuelaNature2006} or spin waves \cite{KajiwaraNature2010}. Thus this demonstration of strong magnon-photon coupling in planar thin-film devices provides a crucial stepping stone for the development of more complex quantum information systems.

We acknowledge helpful discussions with Andy Kent and Xufeng Zhang. This Work was supported by the U.S. National Science Foundation under Grant No. DMR-1808892 and DOE-Visiting Faculty Program. Work at Argonne, including thin films synthesis, device fabrications, and low-temperature measurements, were supported by the U.S. Department of Energy (DOE), Office of Science, Materials Science and Engineering Division. The use of Center for Nanoscale Materials for lithographic processing is supported by DOE-BES, under Contract No. DE-AC02-06CH11357.

\bibliography{ref,ref_QIS}

\newpage

\onecolumngrid

\newpage

\LARGE{\textbf{Supplemental Materials:}}
\section{Strong magnon-photon coupling in ferromagnet-superconductor thin-film devices}
\normalsize
\textit{by} Yi Li, et al.
\newline

\renewcommand{\theequation}{S-\arabic{equation}}
\setcounter{equation}{0}  
\renewcommand{\thefigure}{S-\arabic{figure}}
\setcounter{figure}{0}  

\section{1. Superconductivity transition}

\begin{figure}[htb]
 \centering
 \includegraphics[width=3.0in]{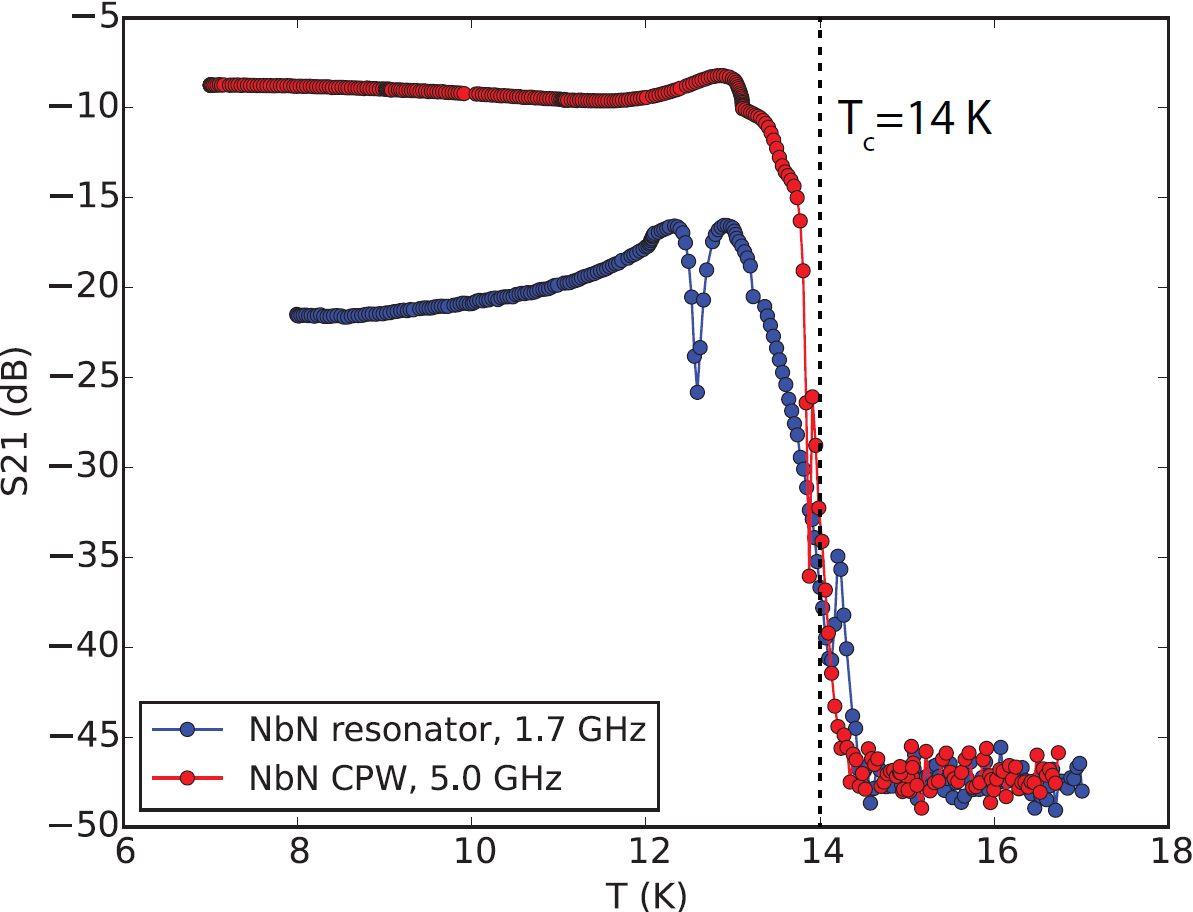}
 \caption{S-parameter for a NbN coplanar waveguide (5.0 GHz) and a NbN resonator (1.7 GHz), as a function of temperature. }
 \label{figS1}
\end{figure}

In Fig. \ref{figS1}, we show the microwave transmission for a NbN coplanar waveguide (CPW) and a NbN resonator as a function of temperature, both unloaded. Here the CPW is for the ferromagnetic resonance measurement of Py stripe. Both circuits demonstrate a sharp transition at 14 K, where $S_{21}$ are reduced by 30 to 40 dB down to the background noise level of -47 dB. Taking account of the 33 dB amplifier and a 5 dBm power input, we obtain a noise background power of -75 dBm in our circuit.

\section{2. Ferromagnetic resoance of Py stripe}

\begin{figure}[htb]
 \centering
 \includegraphics[width=6.0 in]{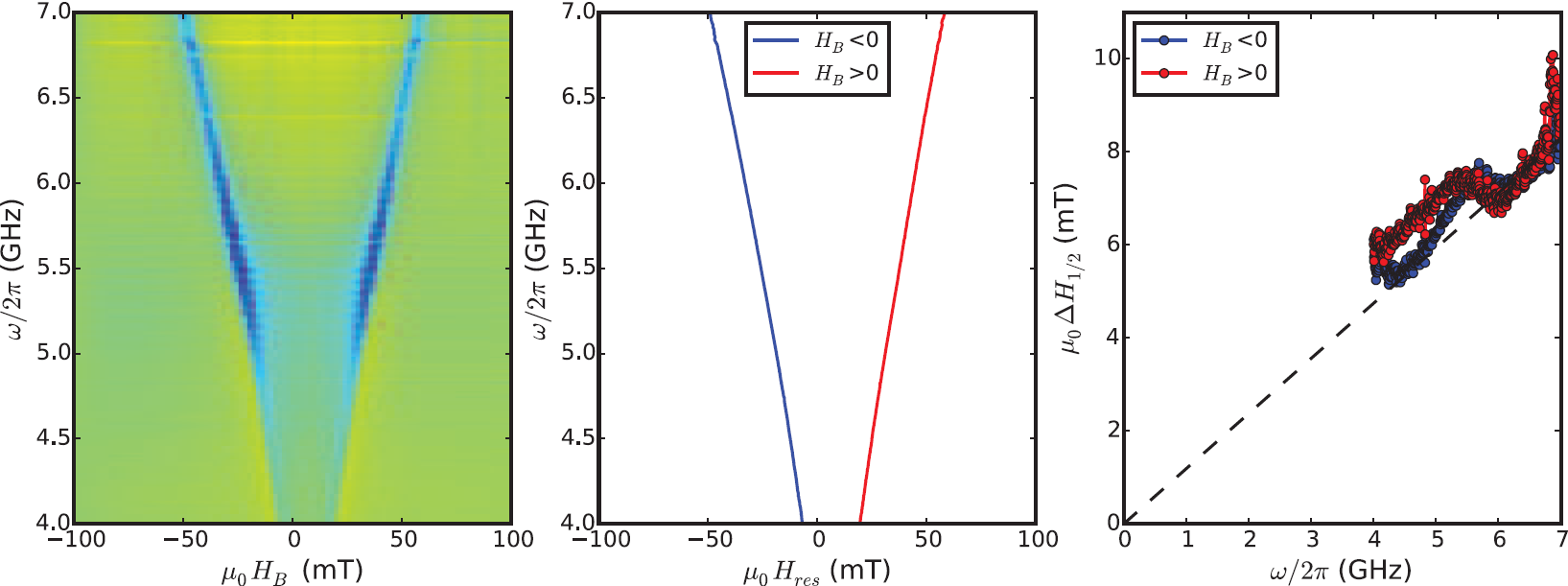}
 \caption{(a) FMR spectra of Py(30 nm) with $L=900$ $\mu$m, measured at 1.4 K and $\theta=0^\circ$. (b-c) Extracted (b) resonance field and (c) linewidth from (a). The dashed line in (c) is a linear fit with a Gilbert damping of $\alpha=0.017$.}
 \label{figS2}
\end{figure}

Fig. \ref{figS2} shows the extracted resonance field $\mu_0 H_{res}$ and linewidth $\mu_0 H_{1/2}$ for the Py(30 nm) stripe, which are used to locate the magnon modes and obtain the magnon damping rate. In Fig. \ref{figS2}(c) the linewidths can be fitted to a Gilbert-type linewidth broadening, $\mu_0 H_{1/2} = \mu_0 H_0 + 2\alpha\omega/\gamma$, which yields $\alpha=0.017$. This value is larger than the typical damping of Py ($\sim 0.007$) at room temperature, owing to the low-temperature damping enhancement. At 5 GHz the linewidth is 6 mT, which are used in the main text to calculate the magnon damping rate.

\section{3. Peak positions and linewidths of magnon-photon hybrid modes}

In this section we will show the exact equations that we use to analyze the hybrid-mode peak positions and linewidths in the main text. For the magnon-photon hybrid modes as described in Eq. (1) of the main text, the eigenmodes come from the following equation:

\begin{equation}\label{eqS01}
    [i(\omega_p-\omega)+\kappa_p][i(\omega_m-\omega)+\kappa_m] = g^2
\end{equation}
with two complex solutions as the magnon-photon hybrid mode ($\omega_{mp}$):

\begin{equation}\label{eqS02}
\tilde{\omega}_{mp}^\pm = {\omega_p+\omega_m \over 2}-i{\kappa_p+\kappa_m \over 2} \pm \sqrt{\left( {\omega_p-\omega_m \over 2}-i{\kappa_p-\kappa_m \over 2} \right)^2+g^2}
\end{equation}

The real and imaginary parts of Eq. (\ref{eqS02}) denote the peak position and linewidth of the hybrid mode. In the low-damping limit, the solution can be expressed as:
\begin{equation}\label{eqS03}
\omega_{mp}^\pm = {\omega_p+\omega_m \over 2} \pm {\sqrt{\left( {\omega_p-\omega_m} \right)^2+4g^2} \over 2}
\end{equation}
\begin{equation}\label{eqS04}
\Delta\omega_{mp}^\pm = {\kappa_p+\kappa_m \over 2} \pm {\kappa_p-\kappa_m \over 2}\times {\omega_p-\omega_m \over \sqrt{\left( \omega_p-\omega_m \right)^2+4g^2}}
\end{equation}

In the main text, we use Eqs. (\ref{eqS03}) and (\ref{eqS04}) to fit the data and extract the coupling strength. We note that also the magnon damping rate is comparable to the coupling strength, the exact solution of Eq. (\ref{eqS02}) can be still well-approximated by Eqs. (\ref{eqS03}) and (\ref{eqS04}).


\section{4. Magnetization state measured by anisotropic magnetoresistance}

\begin{figure}[htb]
 \centering
 \includegraphics[width=5.0 in]{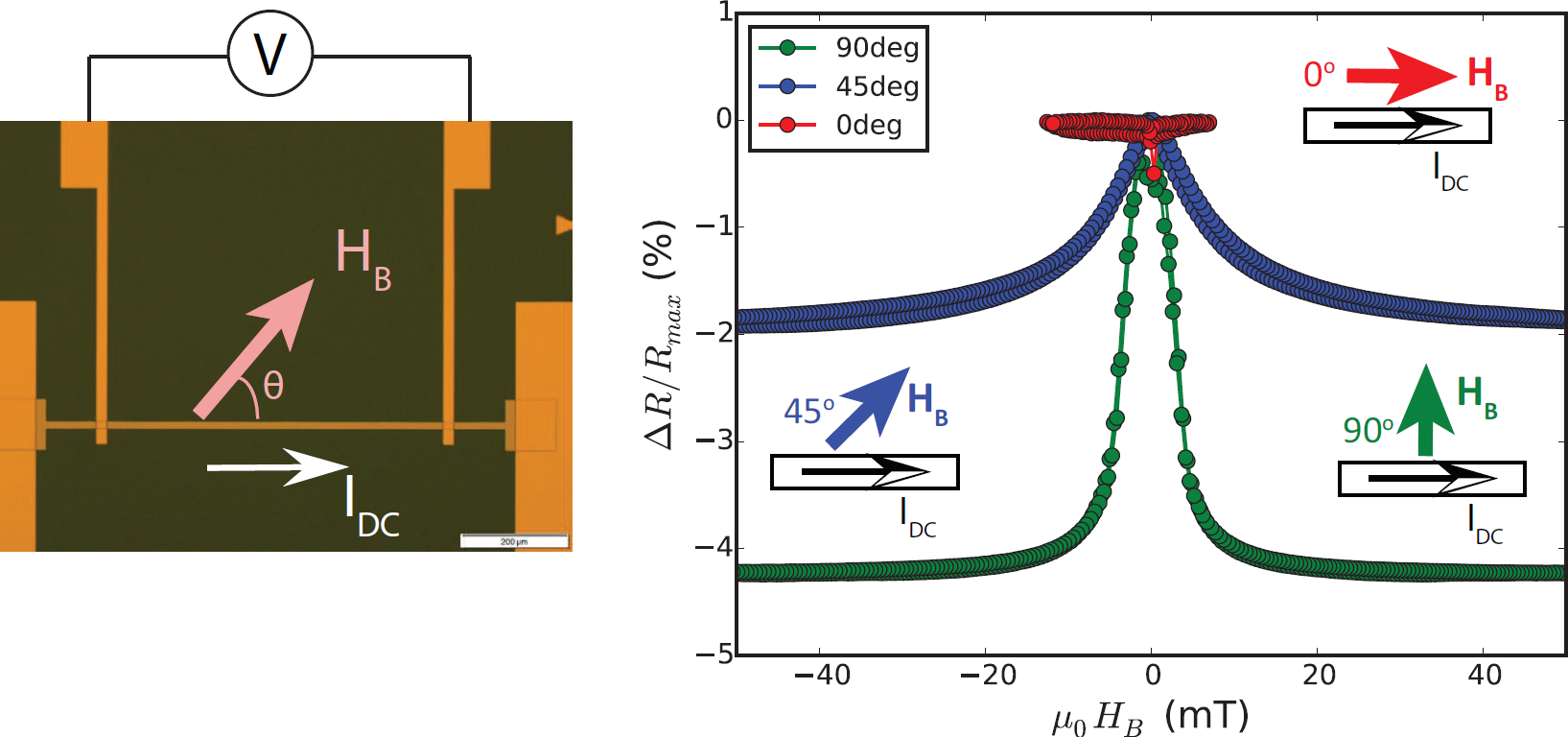}
 \caption{Anisotropic magnetoresistance of a Py(30 nm) stripe with $L=900$ $\mu$m, for different $\theta$.}
 \label{figS3}
\end{figure}

In order to know the magnetization state at low biasing field due to the influence of shape anisotropy, we have conducted anisotropic magnetoresistance (AMR) measurements, using a cryogenic probe station, on a Py(30 nm) stripe with the same deposition and fabrication design. A four-point measurement technique is applied to exclude the contact resistance, with the current direction parallel to the longest edge of the Py stripe.

Fig. \ref{figS3} show the renormalized resistance for $\theta=0^\circ$, $45^\circ$ and $90^\circ$. For $90^\circ$, the resistance become maximal at $\mu_0H_B=0$ mT when the magnetization is along the easy axis and parallel to the current direction. At large biasing field, the magnetization is saturated along the field direction, orthogonal to the current direction and the resistance become minimal. An AMR ratio of 4.2 \% is obtained for Py. The saturation of the magnetization occurs around $\mu_0 H_B=\pm 10$ mT, which is close to that obtained from the spectral gap in Fig. 3(d) of the main text.

For $45^\circ$, because the maximal angle between the magnetization and the current is $45^\circ$, the minimal resistance is located at a higher level than in the case of $\theta=90^\circ$. For $0^\circ$, because it is the easy axis of the Py stripe, the resistance maintains at the maximal value, except around $\mu_0H_B=0$ mT where there is only a small kink during the magnetization switching.

\section{5. Microwave transmission spectra of magnon-photon hybrid system in the nonlinear regime}

\begin{figure}[htb]
 \centering
 \includegraphics[width=5.0 in]{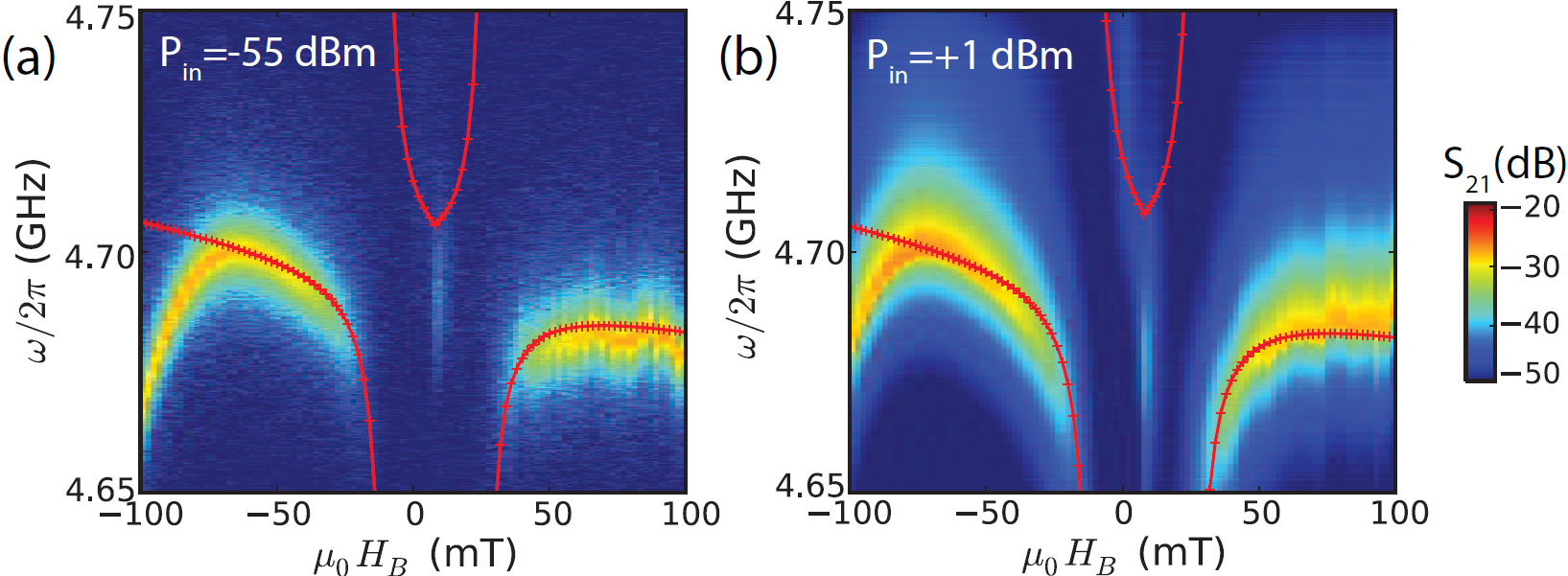}
 \caption{Microwave transmission spectra of a superconducting resonator loaded with a Py(50 nm) strip with $L=900$ $\mu$m. (a) $P_{in}=-55$ dBm. (b) $P_{in}=1$ dBm.}
 \label{figS4}
\end{figure}

Fig. \ref{figS4} compares the power transmission spectra of a superconducting resonator loaded with a Py(50 nm) strip between small ($P_{in}=-55$ dBm) and large ($P_{in}=1$ dBm) microwave inputs. For both cases, The peak positions are determined by the maximal position of the spectra at each biasing field. The two red curves show the best fit to the mode with $g/2\pi=0.139$ GHz for Fig. \ref{figS4}(a) and 0.158 GHz for Fig. \ref{figS4}(b). The extracted peak positions and linewidths are shown in Fig. 4(c) of the main text.

\end{document}